# The Real Dirac Equation

Sokol Andoni*

Technical University of Denmark, Dept. of Chemistry, Kgs. Lyngby 2800, Denmark

*Corresponding author, sond4p@gmail.com or sond@kemi.dtu.dk

**Abstract.** Dirac's leaping insight that the normalized anticommutator of the $\gamma^\mu$ matrices must equal the timespace signature $\eta^{\mu\nu}$ was decisive for the success of his equation. The $\gamma^\mu$-s are *the same* in all Lorentz frames and "describe some new degrees of freedom, belonging to some internal motion in the electron". Therefore, the imposed link to $\eta^{\mu\nu}$ constitutes a separate *postulate* of Dirac's theory. I derive a *manifestly* covariant first order equation from the *direct* quantization of the classical 4-momentum *vector* using the formalism of Geometric Algebra. All properties of the Dirac electron & positron follow from the equation – preconceived 'internal degrees of freedom', *ad hoc* imposed signature and matrices unneeded. In the novel scheme, the Dirac operator is frame-free and manifestly Lorentz invariant. Relative to a Lorentz frame, the classical spacetime frame vectors $e^\mu$ appear instead of the $\gamma^\mu$ matrices. Axial frame vectors (without cross product) of the 3D orientation space defining spin and rotations appear instead of the Pauli matrices; polar frame vectors of the 3D position space naturally define boosts, etc. Not the least, the formalism shows a significantly higher computational efficiency compared to matrices.

1. **Introduction**

Dirac's genial realization [1-3] that the $4 \times 4$ matrices $\gamma^\mu$ have to relate to the timespace signature $\eta^{\mu\nu}$ by

$$\{\gamma^\mu, \gamma^\nu\} \equiv \gamma^\mu\gamma^\nu + \gamma^\nu\gamma^\mu = 2\eta^{\mu\nu} \quad (\mu, \nu = 0,1,2,3), \tag{1}$$

was the key to success for his equation, the Dirac Equation [1-4], DE ($c = 1$ throughout this contribution):

$$(\gamma^\mu \hat{p}_\mu - m)\psi = 0 \text{ (sum over } \mu\text{)}; \quad \hat{p}_\mu = i\hbar\partial_\mu \equiv i\hbar\,\partial/\partial x^\mu. \tag{2}$$

The set of $\gamma^\mu$ satisfying (1) makes part of the Dirac algebra – a special case of Clifford algebras [3-5]. The algebra of $4 \times 4$ complex matrices has an equivalent real dimension of 32. In the standard formulation of DE the $\gamma^\mu$-s are a fundamental representation of some *internal degrees of freedom* of the electron [1-4] and the *same* matrices appear at different Lorentz frames. Therefore, the call for the timespace signature $\eta^{\mu\nu}$ from Special Relativity (SR) in (1) corresponds to a separate postulate of Dirac's theory.

Now, in SR the scalar product of orthonormal Lorentz frame vectors $\{e^\mu\}$ defines the signature:

$$e^\mu \cdot e^\nu \equiv \eta^{\mu\nu} = \tfrac{1}{2}\{e^\mu, e^\nu\}; \quad \text{we choose here the signature (1,3), i.e.} \quad \eta^{\mu\nu} = (+---)\delta^{\mu\nu}. \tag{3}$$

The *conjecture* we are going to prove in the following is that the formal substitution of the *matrices* $\gamma^\mu$ in the standard DE (2) by the frame *vectors* $e^\mu$ from (3) yields the physically best founded relativistic first order equation. This corresponds to the *direct quantization* of the 4-momentum p of modulus $m$ (both Lorentz invariant) with inborn SR signature as shown in (3). In addition to replacing matrices with vectors, we will also make an unnecessary but esthetically rewarding choice, namely choose reals $\mathbb{R}$ as the scalar field of our 32D algebra. In the emerging formalism, the pseudoscalar and all the rest of the *complex structure* surge from geometric objects – multivectors squaring to $-1$, as explained after Eq. (6) below and in Section 2.

The second equality in Eq. (3) tells us that the scalar product is the symmetric part (the anticommutator) of the Clifford, or geometric *vector* product, thus rendering explicit the algebra in (1). The antisymmetric part – the commutator, is Grassmann's wedge product ($\wedge$) [6], so that the geometric product $e^\mu e^\nu$ of two orthonormal frame vectors takes the form (below $\{e^\mu, e^\nu\} = e^\mu e^\nu + e^\nu e^\mu$; $[e^\mu, e^\nu] = e^\mu e^\nu - e^\nu e^\mu$):

$$e^\mu e^\nu \equiv e^\mu \cdot e^\nu + e^\mu \wedge e^\nu = \eta^{\mu\nu} + e^\mu \wedge e^\nu; \quad e^\mu \cdot e^\nu = \tfrac{1}{2}\{e^\mu, e^\nu\}; \quad e^\mu \wedge e^\nu = \tfrac{1}{2}[e^\mu, e^\nu]. \tag{4}$$

For $\lambda \neq \mu \neq \nu \neq \lambda$ the *bivector* $e^\mu e^\nu \equiv e^{\mu\nu} = e^\mu \wedge e^\nu$ (the *trivector* $e^{\lambda\mu\nu} = e^\lambda \wedge e^\mu \wedge e^\nu$) defines an oriented area (resp. volume) element in spacetime. The *tetravector* $e^{0123}$ (an oriented 4-volume element) is the frame multivector of highest grade in spacetime. In general, for any three spacetime vectors u, v, w the geometric product is associative and distributive:

$$uv = u \cdot v + u \wedge v; \quad u \cdot v = \tfrac{1}{2}\{u, v\}; \quad u \wedge v = \tfrac{1}{2}[u, v]; \quad (uv)w = u(vw) = uvw;$$

$$u(v + w) = uv + uw; \quad a(v + w) = av + aw; \quad a \in \mathbb{R} \tag{5}$$

Relations (4, 5) define the spacetime algebra, STA of Hestenes [7, 8] – a real 16D Clifford algebra $\mathcal{Cl}_{(1,3)}$ ((1,3) stands for the signature in Eq. (3)) generated by the action of the geometric product onto the 4-vectors. Hestenes also proposed an STA DE [6, 7] without matrices and with the complex structure arising from the STA (multi)vectors alone. However, spin (the electron's 'internal degrees of freedom') has been put by hand in the equation, thereby diminishing its predictive power and symmetry in comparison to the standard DE.

Now, returning to the conjecture formulated after Eq. (3) above, I proposed recently the most direct form of first order relativistic equation obtained by quantizing the classical 4-momentum *vector* p of an electron [9]; see Eqs. (12, 13) in Section 2. The 4-vector p is relativistic invariant with modulus equal to the rest mass $m$ of the electron – also relativistic invariant. One can apply the quantization postulate $Q$ directly to p, leading to the manifestly covariant Equation (notice, the standard DE is not manifestly covariant):

$$Q: \{p = p_\mu e^\mu; |p| = m\} \rightarrow \{\hat{p}\psi = m\psi; \hat{p} \equiv i\hbar\nabla = i\hbar e^\mu \partial_\mu\}. \quad \text{From Special Relativity: } e^\mu \cdot e^\nu = \eta^{\mu\nu}. \quad (6)$$

The imaginary unit $i$ entering with the momentum operator expands the scalar field of STA from real $\mathbb{R}$ to complex numbers $\mathbb{C}$ after quantization. From Eq. (6) one can certainly build a scheme *equivalent* to the one I will present below. However, with STA on $\mathbb{C}$ the complex structures generated from on one side the multivectors and on the other side from the algebraic $i$, would *mix*, complicating computations and rendering the structure unappealing (the algebraic $i$ fits better to matrices). We will amend this by expanding ST to a 5D *real* vector space, with a complex structure arising only from the multivectors of the respective 32D (Clifford) geometric algebra. The fifth dimension turns out to embody *reflection/ handedness* that becomes in the proposed scheme as fundamental as the four dimensions of *spacetime*. We will see shortly how the *real* 5D spacetime-reflection, STR, accommodates the quantization postulate (see Eq. (12)).

After a quick presentation of STR and STR DE, I will derive standard results such as symmetries, currents, spin and nonrelativistic approximation. Few supporting relations appear in the Appendix [10].

2. **Swift presentation of STR and STR DE**

The 5D STR real vector space comprises a Hermitian frame vector $e^5$ in addition to the four timespace frame vectors $\{e^\mu\}$. It plays a similar role in STR DE as the Dirac $\gamma^5$ matrix in the standard DE, therefore the index. The quintet of frame vectors $\{e^\mu, e^5\}$ under the action of the Clifford product generates the real 32D $\mathcal{Cl}_{(2,3)}$ (signature (2,3)) algebra X of STR with the following basis expressed in terms of the frame vectors:

$$X_{basis}: \{1, e^\lambda, e^{\lambda\mu}, e^{\lambda\mu\nu}, e^{0123}, e^5, e^{\lambda 5}, e^{\lambda\mu 5}, e^{\lambda\mu\nu 5}, e^{01235}\}; \quad \lambda, \mu, \nu = 0,1,2,3; \quad \lambda \neq \mu \neq \nu \neq \lambda. \quad (7)$$

Geometric product of STR basis vectors: $\quad e^\tau e^\upsilon \equiv e^{\tau\upsilon} = e^\tau \cdot e^\upsilon + e^\tau \wedge e^\upsilon = \zeta^{\tau\upsilon} + e^\tau \wedge e^\upsilon;$

The *signature* of STR is (2, 3), i.e.: $\quad \zeta^{\tau\upsilon} \equiv e^\tau \cdot e^\upsilon = (+ - - - +)\delta^{\tau\upsilon}; \quad \tau, \upsilon = 0,1,2,3,5. \quad (8)$

Upright letters stand for (multi)vectors; italic letters stand for scalars. The Hermite conjugate † [2] of an element A ∈ X combines the parity transformation (see Eq. (21)) $e^0 A e^0$, reversal $\widetilde{A}$ (corresponding to matrix transpose), and a factor of $-1$ for each $e^5$ in a multivector, e.g. $(e^{105})^\dagger = -e^0 \widetilde{e^{105}} e^0 = -e^0 e^{501} e^0 = e^{105}$.

This form of † does not 'send' the conjugate to the reciprocal basis. 16 elements of the basis in (7) square to $-1$, allowing for a rich complex structure in X. Of these only the pentavector $e^{01235}$ is both isotropic (i.e. it does not favor any ST frame vector) and commutes with all elements of X; it is the *geometric pseudoscalar* İ:

$$\dot{I} \equiv e^{01235}; \quad \dot{I} = \tilde{\dot{I}}; \quad \dot{I}^2 = -1; \quad \dot{I}^\dagger = -e^0 e^{01235} e^0 = -\dot{I}; \quad \dot{I}e^\tau = e^\tau \dot{I}; \quad \tau = 0,1,2,3,5. \tag{9}$$

The $X_{basis}$ in (7) can now be expressed succinctly as ($\langle ab \rangle_0$ below extracts the *scalar* part of ab):

$$\{1, e^\tau, e^{\tau\upsilon}, \dot{I}e^{\tau\upsilon}, \dot{I}e^\tau, \dot{I}\}; \quad \tau \neq \upsilon; \quad \tau,\upsilon = 0,1,2,3,5. \quad \text{Orthogonality: } \{a, b \in X_{basis}; a \neq b\} \Rightarrow \langle ab \rangle_0 = 0 \tag{10}$$

A generic element $A \in X$ can be expressed in terms of the basis (10) as (Einstein's summation active):

$$X \ni A = \left(a_{(0)} + a_{(5)}\dot{I}\right) + \left(a_{(1)\tau} + a_{(4)\tau}\dot{I}\right)e^\tau + \left(a_{(2)\tau\upsilon} + a_{(3)\tau\upsilon}\dot{I}\right)e^{\tau\upsilon}; \quad a_{(\omega)..} \in \mathbb{R}; \quad \omega, \tau, \upsilon = 0,1,2,3,5. \tag{11}$$

We can lift indices up and down, i.e. swap between reciprocal bases in X, by the appropriate form of the signature in (8). Finally, the **quantization postulate** $Q_{STR}$ yields the **STR DE** ($m$ – rest mass of the electron):

$$Q_{STR}: \{p = p_\mu e^\mu; |p| = m\} \to \{(\hat{p} - m)\psi = 0; \quad \hat{p} = \hbar\dot{I}\nabla = \hbar\dot{I}e^\mu\partial_\mu; \quad \partial_\mu \equiv \partial/\partial x^\mu = \eta_{\mu\nu}\partial/\partial x_\nu\}. \tag{12}$$

Then the STR DE minimally coupled to an external electromagnetic potential $A = e^\mu A_\mu$ becomes:

$$(\hat{P} - m)\psi = 0 \quad \text{with} \quad \hat{P} = \hbar\dot{I}\nabla + eA = e^\mu(\hbar\dot{I}\partial_\mu + eA_\mu); \quad e - \text{charge of the electron}. \tag{13}$$

By inspection, the operator in (12) comprises tetravectors $\dot{I}e^\mu$ sharing $e^5$ and a scalar ($m$). Therefore, for the equation to make sense, the free field $\psi$ *must* comprise terms of all orders in X, in particular the vector $e^5$. We will see in the following that this is indeed the case. But before trying to define the form of the field $\psi$ in STR, it is useful to present three subspaces of X ($j, k = 1,2,3$) (alternative generators can be written down):

**X** with basis: $\{1, \mathbf{x}_j = \mathbf{x}^j \equiv e^{j0}, \mathbf{x}_{jk} \equiv \mathbf{x}_j\mathbf{x}_k, \mathbf{x}_{123} = e^{0123} = \dot{I}e^5\}$; generators: $\{\mathbf{x}_j\}$; 8D

**Σ** with basis: $\{1, \boldsymbol{\sigma}_j = \boldsymbol{\sigma}^j \equiv e^{j05}, \boldsymbol{\sigma}_{jk} = \mathbf{x}_{jk} = \epsilon_{jkl}\dot{I}\boldsymbol{\sigma}_l, \boldsymbol{\sigma}_{123} = \dot{I}\}$; generators: $\{\boldsymbol{\sigma}_j\}$; 8D

**Ξ** with basis: $\{1, e^5, \mathbf{x}_j, \boldsymbol{\sigma}_j, \dot{I}\mathbf{x}_j, \dot{I}\boldsymbol{\sigma}_j, \dot{I}e^5, \dot{I}\}$; **Ξ = XΣ**; generators: $\{e^5, \boldsymbol{\sigma}_j\}, \{e^5, \mathbf{x}_j\}$, or $\{\dot{I}, \mathbf{x}_j\}$; 16D. (14)

It is clear that **Ξ** is isomorphic to $\{\mathbf{X}, e^5\mathbf{X}\}$, or $\{\boldsymbol{\Sigma}, e^5\boldsymbol{\Sigma}\}$, or $\{\mathbf{X}, \dot{I}\mathbf{X}\}$ (with, e.g. $e^5\mathbf{X}: \{e^5 a \mid a \in \mathbf{X}\}$). Also, X is isomorphic to $\{\Xi, e^0\Xi\}$, a relation that will guide our choice of a form for the spinor $\psi$ in Sec. 3 (see (23)).

The 3D vectors $\mathbf{x}_j, \boldsymbol{\sigma}_j$ are Hermitian. Since $e^0 \mathbf{x}_j e^0 = -\mathbf{x}_j$ (parity-odd) and $e^0 \boldsymbol{\sigma}_j e^0 = \boldsymbol{\sigma}_j$ (parity-even), $\mathbf{x}_j$ are *polar* vectors and as we will see shortly generate boosts, while $\boldsymbol{\sigma}_j$ are *axial* vectors – generators of spin, e.g. $\hbar \boldsymbol{\sigma}_j/2$, and rotors (see $\mathbf{J}_j$ below). More precisely, the ST *bivectors* $e^\mu \wedge e^\nu$ – independent of $e^5$, generate the Lorentz group: $\mathbf{J}_l \equiv \epsilon_{jkl} e^{jk} = -\mathbf{i}\boldsymbol{\sigma}_l$ for rotors and $\mathbf{K}_j \equiv e^{j0} = \mathbf{x}_j$ for boosts. The basic Lorentz operator is obtained by exponentiation of $\mathbf{J}_j, \mathbf{K}_j$ (e.g., $e^{-\mathbf{i}\boldsymbol{\sigma}_2 \vartheta/2} = \cos\frac{\vartheta}{2} - \mathbf{i}\boldsymbol{\sigma}_2 \sin\frac{\vartheta}{2}$; $e^{\mathbf{x}_1 \alpha/2} = \cosh\frac{\alpha}{2} + \mathbf{x}_1 \sinh\frac{\alpha}{2}$):

$$S = e^{\mathbf{J}_j \vartheta_j/2 + \mathbf{K}_j \alpha_j/2}, \quad \text{where:} \quad \begin{cases} \mathbf{J}_j = -\mathbf{i}\boldsymbol{\sigma}_j \text{ rotor part;} & \vartheta_j - \text{ Euclidean rotation angles} \\ \mathbf{K}_j = \mathbf{x}_j \text{ boost part;} & \alpha_j - \text{ rapidity, hyperbolic angles} \end{cases};$$

$$S^\dagger = e^0 \tilde{S} e^0 = e^0 S^{-1} e^0 \qquad \text{(expressed in the 'unprimed' frame).} \tag{15}$$

We will show shortly that frame vectors transform by *two*-sided operations $e^\mu \to Se^\mu \tilde{S}$, while Pauli spinors (see Eqs. (23, 25)) by *one*-sided operations $\varphi \to S\varphi$ with the distinctive *half angles* for spinors. As an aside, $\mathbf{J}_j$ and $\mathbf{K}_j$ define the frame for the Killing vector appearing in the STR commutator [x, p]; see (A1) in [10].

### 3. Lorentz transformation of STR DE and the form of the STR Dirac spinor

From (15) $\tilde{S} = S^{-1} \ne S^\dagger$, i.e. S is not unitary; this is expected due to the opposite behavior of the rotor and the boost generators under Hermite conjugation: $(\mathbf{i}\boldsymbol{\sigma}_j)^\dagger = -\mathbf{i}\boldsymbol{\sigma}_j$; $\mathbf{x}_j^\dagger = \mathbf{x}_j$. S is the Lorentz operator (two-sided) for the *frame vectors*, with $\mathcal{L}$ below standing for Lorentz transformation (notice that $e^\mu \cdot e_\nu = \delta^\mu_\nu$):

$$\mathcal{L}: e^\mu \to e'^\mu = Se^\mu \tilde{S}; \quad \mathcal{L}: e_\mu \to Se_\mu \tilde{S} = e'_\mu. \quad \text{Obviously: } Se^5 \tilde{S} = e^5 \text{ and } S\mathbf{i}\tilde{S} = \mathbf{i}. \tag{16}$$

Now, the operator $p - m$ (for simplicity we drop the hat from $\hat{p}$) in the STR DE (12) is Lorentz invariant:

$$\mathcal{L}: p - m \to p' - m = p - m; \quad p = \hbar\mathbf{i}\nabla = \hbar\mathbf{i}e^\mu \partial_\mu = p' = \hbar\mathbf{i}\nabla' = \hbar\mathbf{i}e'^\mu \partial'_\mu. \tag{17}$$

From (16, 17) the component $\partial_\mu$ of the Lorentz invariant 4-vector operator $\nabla$ is the projection $e_\mu \cdot \nabla$, i.e.:

$$\text{from: } \nabla = e^\mu \partial_\mu = \nabla' = e'^\mu \partial'_\mu \Rightarrow \mathcal{L}: \{\partial_\mu = e_\mu \cdot \nabla\} \to \{\partial'_\mu = e'_\mu \cdot \nabla = Se_\mu \tilde{S} \cdot \nabla\}. \tag{18}$$

Relations (16-18) define the *general* Lorentz operator $\mathcal{S}$: (below $p_\mu = \hbar\mathbf{i}\partial_\mu$; see also Eq. (5)):

$$\mathcal{L}: p \to p' = \mathcal{S}p\mathcal{S}^{-1} = \mathcal{S}p\tilde{\mathcal{S}} = p. \text{ In components: } \mathcal{S}e^\mu p_\mu \tilde{\mathcal{S}} = \mathcal{S}e^\mu \tilde{\mathcal{S}} \mathcal{S} p_\mu \tilde{\mathcal{S}} = \mathbf{i}\hbar e'^\mu \mathcal{S}(e_\mu \cdot \nabla)\tilde{\mathcal{S}} =$$

$$\tfrac{1}{2}\mathbf{i}\hbar e'^\mu \mathcal{S}(e_\mu \nabla + \nabla e_\mu)\tilde{\mathcal{S}} = \tfrac{1}{2}\mathbf{i}\hbar e'^\mu (e'_\mu \nabla + \nabla e'_\mu) = \mathbf{i}\hbar e'^\mu (e'_\mu \cdot \nabla) = e'^\mu p'_\mu = \mathbf{i}\hbar Se^\mu (Se_\mu \tilde{S} \cdot e^\nu \partial_\nu)\tilde{S}. \tag{19}$$

Taking $\mathcal{S}^{-1} = \tilde{\mathcal{S}}$ follows from $p = \tilde{p}$ and $\tilde{S} = S^{-1}$. $\mathcal{S}$ commutes with Lorentz-invariant (multi)vectors, like p, $\nabla$ or the 4-position $x(= e^\mu x_\mu = e_\mu x^\mu)$. The last expression in (19) states $\mathcal{S}$ in terms of S; however, clearly $\mathcal{S} \ne S$ as $\mathcal{S}p\tilde{\mathcal{S}} = p' = p \ne Sp\tilde{S} = p_\nu e'^\nu$. Not distinguishing $\mathcal{S}$ from S has led to an incorrect treatment of the relativistic covariance in [9]. We will use in the following the Lorentz operator $\mathcal{S}$ in the general discussion of the symmetries of STR DE and of the Dirac field ψ. $\mathcal{S}$ can always be expressed through S as in the last expression of (19). Now, the STR DE transforms as:

$$\mathcal{L}: \{(p - m)\psi = 0\} \to \{\mathcal{S}(p - m)\psi = (p' - m)\psi' = 0\}; \qquad \psi' \equiv \mathcal{S}\psi. \tag{20}$$

As in the standard case [3, 4], the relativistic covariance in (20) means form-invariance for the STR DE in the primed and in the original frames. However, the STR DE and the standard DE are quite different. The frame-free operator in (20) has a clear physical meaning and is Lorentz invariant, making STR DE *manifestly* covariant; the (classical) frame vectors naturally follow the algebra in (3). On the other side the standard DE – (*not* manifestly) covariant, uses the *same* $\gamma^\mu$ matrices at different Lorentz frames, vaguely representing 'some internal degrees of freedom of the electron' and assumed *ad hoc* to follow the algebra in (1).

Now, how do we express the spinor transformation $\psi \to \mathcal{S}\psi$ from (20) in terms of S? $\mathcal{S}$ reduces to S in Lorentz transformations of e.g. the STR Pauli spinors we will meet in Eq. (23). But we cannot substitute $\mathcal{S}$ by S in general, as can be seen in at least two ways. First, applying the last expression of (19) to (20) yields:

$$\mathcal{S}(p - m)\psi = (p' - m)\mathcal{S}\psi = S[\hbar i e^\mu (S e_\mu \tilde{S} \cdot \nabla) - m]\tilde{S}S\psi = 0. \tag{21}$$

For $\mathcal{S} = S$ we would obtain a one-sided S operator; yet, the expression in the square brackets *is not* the Dirac operator. A more direct way appears in the Appendix [10], showing the Lorentz transformation of the free field solutions $\psi_\pm$ in Eqs. (A15-17). These arguments lead to the conclusion that S is the *one-sided* STR Lorentz operator on the 'genuine' spinor constituents of ψ, like the Pauli spinors. However, we need $\mathcal{S}$ in order to transform correctly the Dirac spinor as a whole, as it also comprises scalar components of position and momentum, see (A5, A9, A11, A14). These components transform in exactly the same way as in (19). Notice at last that from (20) the STR DE as a whole transforms under the general Lorentz operator $\mathcal{S}$ in the same way as the Dirac spinor ψ. The two-sided and one-sided transformations in Eqs. (16), (20) apply also in the case of the discrete symmetries discussed in Sec. 4. For example, the parity transformation of STR DE is:

$$\mathcal{P}: e^\mu \to e^0 e^\mu e^0; \quad \mathcal{P}: (p - m)\psi = 0 \to e^0(p - m)\psi = (p_\mathcal{P} - m)\psi_\mathcal{P} = 0; \quad p_\mathcal{P} = e^0 p e^0; \quad \psi_\mathcal{P} = e^0 \psi. \tag{22}$$

We look now at the **form of the STR Dirac spinor** ψ. The Lorentz-transformed spinor $\psi'$ in (20) is also a Dirac spinor and as obvious from (15) it would in general comprise linear combinations of elements of the algebra Ξ with basis (14). Therefore, we expect for the space of spinors {ψ} to at least comprise Ξ; see (14,

A11, A14). Parity would then expand the space of spinors to the whole algebra X (see (11)), $\{\psi\} \subseteq X$. The isomorphism between the algebras $\{\Xi, e^0\Xi\}$ and X allows to isolate the effect of parity by splitting $\psi$ into a parity-even $\varphi$ and a parity-odd $\chi$ STR Pauli spinors with the help of the orthogonal projectors $(1 \pm e^0)$:

$$\psi = \tfrac{1}{2}[(1 + e^0)\psi + (1 - e^0)\psi] = \varphi + \chi; \quad \varphi \equiv \tfrac{1}{2}(1 + e^0)\psi; \quad \chi \equiv \tfrac{1}{2}(1 - e^0)\psi; \quad \varphi, e^5\chi \in \Sigma \bmod e^0. \quad (23)$$

The form of the two spinors ensures $\varphi + \chi \in \Xi$ (see (A11-A14)), as demanded by $\mathcal{L}$ in (20). In (23) $\bmod\ e^0$ reminds us of the projectors inherent in $\varphi, \chi$. Therefore, under parity we can write the *eigen*-equations:

$$\mathcal{P}: \psi \to \psi_\mathcal{P} = e^0\psi = \varphi - \chi, \quad \text{or} \quad e^0\varphi = \varphi; \quad e^0\chi = -\chi \quad \text{(notice that } e^0 e^5 \chi = e^5\chi \in \Sigma\text{)}. \quad (24)$$

Now, as by standard convention, we pick the spin basis along $\boldsymbol{\sigma}_3$; then each of the Pauli spinors split into spin up and spin down by the orthogonal projectors $(1 \pm \boldsymbol{\sigma}_3)$, e.g. in the case of $\varphi$ (below $a, b \in \mathbb{R}$):

$$\varphi = u_\varphi + d_\varphi \in \Sigma; \quad u_\varphi \equiv \tfrac{1}{2}(1 + \boldsymbol{\sigma}_3)\varphi; \quad d_\varphi \equiv \tfrac{1}{2}(1 - \boldsymbol{\sigma}_3)\varphi; \quad u_\varphi, \boldsymbol{\sigma}_1 d_\varphi \in \{a + b\mathfrak{i}\} \bmod \boldsymbol{\sigma}_3. \quad (25)$$

The known *eigen*values $\pm 1$ for spin up / down follow from (25) in the form of the *eigen*-equations below:

$$\boldsymbol{\sigma}_3 \varphi = u_\varphi - d_\varphi, \quad \text{or} \quad \boldsymbol{\sigma}_3 u_\varphi = u_\varphi, \quad \boldsymbol{\sigma}_3 d_\varphi = -d_\varphi; \quad \bmod\ \boldsymbol{\sigma}_3 \text{ reminds us of the projectors in } u_\varphi, d_\varphi. \quad (26)$$

Expressions similar to (25, 26) apply for $e^5\chi = u_\chi + d_\chi$. The form of the spinor in (25) ensures that $\varphi \in \Sigma$. Eq. (A17) [10] illustrates the (one-sided) Lorentz transformation of spinors. STR renders explicit the defining role of $\mathcal{L}$-transformations & space-reversal (parity) on the form of the Dirac field $\psi$ in Eqs. (23, 25). The probability amplitudes for spin up and spin down are proportional to $u_\varphi, \boldsymbol{\sigma}_1 d_\varphi$, respectively. Normalizing, we get the total probability $|u_\varphi|^2 + |\boldsymbol{\sigma}_1 d_\varphi|^2 = 1$. For spin depending on position $u_\varphi, d_\varphi$ are in general functions of position and the condition of normalization above appears as an integral over the 3D space. For spin independent on position (*decoupling* of spin from position, *s-p*), the spatial parts of $u_\varphi$ and $d_\varphi$ equal a common factor $\rho_\varphi$ of modulus 1 (see e.g. (A11-14)), which as in STA [7] renders $\varphi$ proportional to a rotor (below I rewrite Eq. (25) as $\varphi = u_\varphi + d_\varphi = u_\varphi - \mathfrak{i}\boldsymbol{\sigma}_2 d'_\varphi$ with $d'_\varphi \equiv \boldsymbol{\sigma}_1 d_\varphi = \mathfrak{i}\boldsymbol{\sigma}_2 d_\varphi$):

$$\varphi = u_\varphi - \mathfrak{i}\boldsymbol{\sigma}_2 d'_\varphi \stackrel{s-p}{=} \rho_\varphi R_\vartheta = \rho_\varphi e^{-\mathfrak{i}\boldsymbol{\sigma}_2 \vartheta/2} \quad \text{with} \quad \rho_\varphi \cos\tfrac{\vartheta}{2} \equiv u_\varphi; \quad \rho_\varphi \sin\tfrac{\vartheta}{2} \equiv d'_\varphi. \quad (27)$$

We illustrate the working of the spinors in (23-27) with two examples. In the first example to follow, we look at the form of the STR DE in the rest frame of the electron (see also [12]) and in the second example in the Appendix [10] we derive the STR DE free field solutions; see Eqs. (A11-A14).

By plugging $\psi(x) = \int \frac{d^4p}{(2\pi)^4} e^{-ip\cdot x/\hbar}\psi(p)$ into (12) we obtain the STR DE in the momentum space:

$$(p - m)\psi_p = 0, \quad \text{with the 4-momentum vector} \quad p = e^\mu p_\mu. \tag{28}$$

Due to relativistic covariance we can write down the Equation in the rest frame rf of the electron:

$$(e_{rf}^0 - 1)\psi_{rf} = (e_{rf}^0 - 1)(\varphi_{rf} + \chi_{rf}) = 0; \quad me_{rf}^0 = p_{rf} = p. \tag{29}$$

In the rest frame $(e_{rf}^0 - 1)$ is explicitly a parity projector as in (23) (the Dirac operator in a generic frame is obtained by boosting this projector [3]). From (23, 24), $\chi_{rf}$, i.e. the parity-odd part of the spinor vanishes:

$$(e_{rf}^0 - 1)(\varphi_{rf} + \chi_{rf}) = (e_{rf}^0 - 1)\chi_{rf} = -2\chi_{rf} = 0. \tag{30}$$

This confirms what we know about the electron; in the rest frame it has only the two degrees of freedom of spin, represented by the Pauli spinor $\varphi_{rf}$ with explicit expression shown in Eq. (A13).

**4. Conserved currents, Lagrangian, spin magnetic moment and discrete symmetries of STR DE**

These subjects have been reported in detail elsewhere [9] and here I will just touch on them shortly. Similarly to the standard approach we start by taking the Hermite conjugate of the STR DE in (13):

$$(P - m)\psi = [e^\mu(\hbar i\partial_\mu + eA_\mu) - m]\psi \xrightarrow{\dagger} \psi^\dagger\left[e^0 e^\mu e^0 \left(-\hbar i\overleftarrow{\partial}_\mu + eA_\mu\right) - m\right] = \psi^\dagger e^0 \left[e^\mu \left(-\hbar i\overleftarrow{\partial}_\mu + eA_\mu\right) - m\right] e^0 \equiv \overline{\psi}\left[e^\mu\left(-\hbar i\overleftarrow{\partial}_\mu + eA_\mu\right) - m\right] e^0 = 0. \tag{31}$$

After Hermite conjugation, $\dagger$, $\overleftarrow{\partial}_\mu$ act to the left. Right-multiply the last equation by $\overline{\psi}^\dagger = e^0\psi$, left-multiply the STR DE by $\overline{\psi} = \psi^\dagger e^0$ and by subtraction obtain the ***conserved probability current***:

$$(\partial_\mu\overline{\psi})e^\mu\psi + \overline{\psi}e^\mu(\partial_\mu\psi) = \partial_\mu(\overline{\psi}e^\mu\psi) = 0 \quad \text{with probability density} \quad \overline{\psi}e^0\psi = \psi^\dagger\psi \geq 0. \tag{32}$$

See Table 1 for more details on the current components. The Dirac conjugate $\overline{\psi} = \psi^\dagger e^0$ is the Hermite conjugate of the parity transformed $\psi$, i.e.: $(e^0\psi)^\dagger = \overline{\psi}$. It substitutes the $\psi^\dagger$ of the nonrelativistic quantum mechanics (see (34)) and from (20) its Lorentz transform from the perspective of the 'unprimed frame' is:

$$\text{From} \quad \mathcal{L}: \psi^\dagger \to \{(S\psi)^\dagger = \psi'^\dagger = \psi^\dagger e^0 \tilde{S} e^0\} \quad \Rightarrow \quad \mathcal{L}: \{\overline{\psi} = \psi^\dagger e^0\} \to \{\psi'^\dagger e^0 = \overline{\psi}\tilde{S} \equiv \overline{\psi}'\}. \tag{33}$$

Lorentz transformation of three STR Dirac bilinears takes the form (only the spinors transform; e.g., in the case of the currents the equation of conservation is Lorentz form-invariant, $\partial_\mu\overline{\psi}e^\mu\psi = 0 \to \partial'_\mu\overline{\psi}'e'^\mu\psi' = 0$; in $\overline{\psi}e^\mu\psi$ we detach the operator $\partial_\mu$ from the bilinear component, thus 'fixing' the frame vectors $e^\mu$):

$$\mathcal{L}: \{\overline{\psi}\psi \to \overline{\psi}'\psi' = \overline{\psi}\tilde{S}S\psi = \overline{\psi}\psi; \quad \overline{\psi}e^\mu\psi \to \overline{\psi}'e^\mu\psi' = \overline{\psi}\tilde{S}e^\mu S\psi; \quad \overline{\psi}e^5\psi \to \overline{\psi}\tilde{S}e^5 S\psi = \overline{\psi}e^5\psi\}. \tag{34}$$

Table 1 lists the 16 Dirac bilinears with the spinor ψ expressed as in (23, 25). It is clear from (34) and Tab. 1 that $\bar{\psi}\psi, \bar{\psi}e^5\psi, \bar{\psi}e^\mu\psi$ are respectively a relativistic scalar, a pseudoscalar and the component of a vector.

| Bilinear | Standard form | STR form | Expanded form in STR (with ψ from Eq. (23)) |
|---|---|---|---|
| Scalar | $\bar{\psi}\psi$ | $\bar{\psi}\psi$ | $\varphi^\dagger\varphi - \chi^\dagger\chi = u_\varphi^\dagger u_\varphi + d_\varphi^\dagger d_\varphi - (u_\chi^\dagger u_\chi + d_\chi^\dagger d_\chi)$ (a) |
| Conserved 4-current | $\bar{\psi}\gamma^\mu\psi$ | $\bar{\psi}e^\mu\psi$ | $\delta^{\mu 0}(\varphi^\dagger\varphi + \chi^\dagger\chi) - \delta^{\mu j}(\varphi^\dagger \mathbf{x}_j\chi + \chi^\dagger \mathbf{x}_j\varphi)$ |
| Tensor / Bivector | $\bar{\psi}\sigma^{\mu\nu}\psi$ (b) | $\bar{\psi}e^\mu \wedge e^\nu\psi$ | $-\varepsilon(\varphi^\dagger \mathbf{x}_j\chi + \chi^\dagger \mathbf{x}_j\varphi) - \mathrm{i}\delta\varepsilon_{jkl}(\varphi^\dagger\boldsymbol{\sigma}_l\varphi - \chi^\dagger\boldsymbol{\sigma}_l\chi)$ (c) |
| Pseudo (axial) vector | $\bar{\psi}\gamma^\mu\gamma^5\psi$ | $\bar{\psi}e^{\mu 5}\psi$ | $\delta^{\mu 0}e^5(\varphi^\dagger\chi + \chi^\dagger\varphi) + \delta^{\mu j}(\varphi^\dagger\boldsymbol{\sigma}_j\varphi + \chi^\dagger\boldsymbol{\sigma}_j\chi)$ (d) |
| Pseudoscalar | $\bar{\psi}\gamma^5\psi$ | $\bar{\psi}e^5\psi$ | $\varphi^\dagger e^5\chi - \chi^\dagger e^5\varphi$ (e) |

**Table 1.** Expressions for the Dirac bilinears in the standard and STR formalisms. Expanded forms of the STR Dirac bilinears appear in the last column, in terms of the Pauli spinors (23). We develop $\bar{\psi}\psi$ further by applying (25).

(a) From (25) the last expression is a real number, the expectation value $\langle\bar{\psi}\psi\rangle$. In the *spin-position* decoupling regime (27) we get the simple form $\rho_\varphi^2 - \rho_\chi^2$, which makes contact to STA's form $\rho^2 \cos\beta$ with $\rho^2 = \rho_\varphi^2 + \rho_\chi^2$ and $\rho_\varphi^2/\rho^2 = \cos^2\frac{\beta}{2}$. I do not use $\beta$ here.

(b) The standard antisymmetric traceless tensor is defined by the commutator of Dirac matrices multiplied by $i$, $\sigma^{\mu\nu} \equiv \frac{i}{2}[\gamma^\mu, \gamma^\nu]$;

(c) $\varepsilon \equiv (\delta^{\mu 0}\delta^{\nu j} - \delta^{0\nu}\delta^{\mu j})$ and $\delta \equiv \delta^{\mu j}\delta^{\nu k}$;

(d) As anticipated in (14), $\boldsymbol{\sigma}_j$ are axial, therefore they appear naturally here. For $m = 0$ the axial currents $\bar{\psi}e^{\mu 5}\psi$ are conserved.

(e) From (25) in STR $\bar{\psi}e^5\psi = \varphi^\dagger e^5\chi - \chi^\dagger e^5\varphi = u_\varphi^\dagger u_\chi + d_\varphi^\dagger d_\chi - u_\chi^\dagger u_\varphi - d_\chi^\dagger d_\varphi$, which changes sign under Hermite conjugation, as a pseudoscalar should. Under the *s-p* decoupling (27), $\bar{\psi}e^5\psi = \langle R_\varphi^\dagger R_\chi\rangle_0(\rho_\varphi^\dagger\rho_\chi - \rho_\chi^\dagger\rho_\varphi)$, where $\langle R_\varphi^\dagger R_\chi\rangle_0$ is the scalar part of $R_\varphi^\dagger R_\chi$.

***The Dirac Lagrangian in STR.*** The STR DE Lagrangian is similar in form with the standard one:

$$\mathfrak{L} = \bar{\psi}(\mathrm{p} - m)\psi = \bar{\psi}(\mathrm{I}\hbar\nabla - m)\psi. \tag{35}$$

As in the standard case $\bar{\psi}$ and $\psi$ are independent and one can recover STR DE from the field equations, most directly by differentiating $\mathfrak{L}$ with respect to $\partial_\mu\bar{\psi}$ and $\bar{\psi}$ obtaining ($\nabla$ in (35) acts to the right):

$\partial_\mu \frac{\delta\mathfrak{L}}{\delta\,\partial_\mu\bar{\psi}} - \frac{\delta\mathfrak{L}}{\delta\bar{\psi}} = 0 \Rightarrow \frac{\delta\mathfrak{L}}{\delta\bar{\psi}} = (\mathrm{I}\hbar\nabla - m)\psi = 0$. The conservation of probability currents (32) follows from the gauge symmetry of the Lagrangian and Noether's theorem [3, 4]. Gauge symmetry: $\{\psi \to e^{\mathrm{I}\theta}\psi: \mathfrak{L} \to \mathfrak{L}\} \Rightarrow$

$\partial_\mu\bar{\psi}\mathrm{x}^\mu\psi = 0$. The Lagrangian is Lorentz-invariant, as it has to be:

$$\mathcal{L}: \{\mathfrak{L} = \bar{\psi}(\mathrm{p} - m)\psi\} \to \{\mathfrak{L}' = \bar{\psi}\tilde{\mathcal{S}}(\mathrm{p} - m)\mathcal{S}\psi = \mathfrak{L}\}. \tag{36}$$

***The spin 1/2 magnetic angular momentum.*** Taking the square of the STR DE (13):

$\mathrm{P}\psi = m\psi \;\;\Rightarrow\;\; \mathrm{PP}\psi = m^2\psi \;\;\Leftrightarrow\;\; [(\hbar\mathrm{I}\nabla + e\mathrm{A})(\hbar\mathrm{I}\nabla + e\mathrm{A}) - m^2]\psi =$

$[\eta^{\mu\nu}(\hbar\mathrm{I}\partial_\mu + eA_\mu)(\hbar\mathrm{I}\partial_\nu + eA_\nu) - m^2 + e\hbar\mathrm{I}(\nabla \wedge A + A \wedge \nabla)]\psi = \{\mathrm{KG} + e\hbar\mathrm{I}[(\nabla \wedge A)]\}\psi =$

$\{\mathrm{KG} + e\hbar\mathrm{I}[-(\nabla A_0) - (\partial_0\mathbf{A}) + \mathrm{I}e^5(\nabla \times \mathbf{A})]\}\psi = \{\mathrm{KG} + e\hbar\mathrm{I}[\mathbf{E} + \mathrm{I}(\boldsymbol{\sigma}, \mathbf{B})]\}\psi \equiv (\mathrm{KG} + e\hbar\mathrm{I}\mathrm{F})\psi = 0.$ (37)

Brackets in e.g. $(\nabla \wedge A)$ or $(\nabla A_0)$ confine the action of the operator.

$\mathbf{A} = A_j\mathbf{x}_j$, $\mathbf{E} = E_j\mathbf{x}_j$ (vector potential and electric field, polar 3D vectors); $(\boldsymbol{\sigma}, \mathbf{B}) \equiv \sigma_j B_j =$
$e^5\mathbf{B}$ (magnetic field, axial 3D vector); $F \equiv \mathbf{E} + \dot{I}(\boldsymbol{\sigma}, \mathbf{B}) = (\nabla \wedge A)$ (Faraday, bivector). (38)

KG stands for the Klein-Gordon term – the symmetric part of PP, comprising grade 0, 5 components. The term $(\nabla \wedge A) = \mathbf{E} + \dot{I}(\boldsymbol{\sigma}, \mathbf{B})$ ($e^5$-independent) is the *Faraday* F, depicting the relativistic invariant *EM field strength* experienced by the electron, as marked by the prefactor $e\hbar$. F is an antisymmetric tensor in the standard formalism [3, 4]; from (38) it is a 4D bivector in STR [7, 9, 11]. The term $e\hbar\dot{I}F$ distinguishes the squared DE from the KG Equation. It represents the '*internal degrees of freedom*' of the electron – the spin, interacting with the EM field. In the nonrelativistic regime, it leads to the term $(\hbar e/2m)(\boldsymbol{\sigma}, \mathbf{B})$ in the Pauli Hamiltonian [9] (see Eq. (A19) [10]), marking the additional potential energy due to the spin magnetic moment of a slow electron with Dirac's gyromagnetic ratio of 2. The experimental spin gyromagnetic ratio for the electron is a factor ~1.00116 larger than 2, the gap arising from QED corrections, which are beyond the scope of DE [3, 4]. Eq. (37) proves that the magnetic moment of spin springs from the direct quantization of the 4-momentum vector – matrices or preconceived internal degrees of freedom unneeded.

***Symmetries of the STR Dirac Field.*** Below I will show few forms of the basic symmetries for the STR Dirac field ψ; other forms are possible as will be shown elsewhere; see also [9]. Let me first introduce $\kappa_j$-conjugation – the operation of sign-swap for the frame vector $e^j$, facilitating in STR, e.g. the standard antiunitary transformations [13] as illustrated in (42, 43) below (($j$) means no sum over repeated $j$):

$$\kappa_j: e^\tau \to \{\kappa_{(j)}e^\tau\kappa_{(j)} = (1 - 2\delta_{j\tau})e^\tau; \; j = 1,2,3; \; \tau = 0,1,2,3,5\}; \quad \kappa_{(j)}^2 = 1; \quad \kappa_{(j)}\dot{I}\kappa_{(j)} = -\dot{I}. \quad (39)$$

Now, as already mentioned in relation to Lorentz transformations in (20) and parity in (22), symmetry operations act on the overall STR DE as one-sided operations, exactly as on the spinor ψ, i.e.:

***Lorentz.*** $\mathcal{L}: p \to \mathcal{S}p\tilde{\mathcal{S}} = p' = p; \quad \mathcal{S}e^\mu\tilde{\mathcal{S}} = Se^\mu\tilde{S} = e'^\mu; \qquad \mathcal{L}: \psi \to \psi' = \mathcal{S}\psi.$ (40)

***Parity.*** $\mathcal{P}: e^\mu \to e^0 e^\mu e^0 = \eta^{\mu\nu}e^\nu; \qquad \mathcal{P}: \psi \to \psi_\mathcal{P} = e^0\psi.$ (41)

***Time Reversal.*** $\mathcal{T}: x_0 \to -x_0; \qquad \mathcal{T}: \psi \to \psi_\mathcal{T} = \boldsymbol{\sigma}_{(j)}\kappa_{(j)}\psi.$ (42)

***Charge conjugation.*** $\mathcal{C}: e \to -e; \qquad \mathcal{C}: \psi \to \psi_\mathcal{C} = \kappa_{(j)}e^{(j)}\psi = e^{05}\boldsymbol{\sigma}_{(j)}\kappa_{(j)}\psi.$ (43)

***CPT.*** $\mathcal{CPT}: (e \to -e)(e^\mu \to e^0 e^\mu e^0)(x_0 \to -x_0); \qquad \mathcal{CPT}: \psi \to \psi_{\mathcal{CPT}} = e^5\psi.$ (44)

Notice the reflection frame vector $e^5$ in (43, 44)! Within an overall phase factor, it can be replaced by $e^{0123}$. It is straightforward to prove that the STR Dirac Lagrangian (35) is invariant under the action of $\mathcal{CPT}$. In contrast to the standard scheme, where the Pauli matrices $\sigma_1, \sigma_3$ are real while $\sigma_2$ is imaginary, the spin vectors $\boldsymbol{\sigma}_j$ in STR are all real and as shown in (42, 43) one can pick any of them for the transformations $\mathcal{T}, \mathcal{C}$ coupled with the respective conjugation from (39). The charge conjugation operator $\kappa_{(j)}e^{(j)}$ is Lorentz-invariant, i.e. it commutes with the Lorentz operator $\mathcal{S}$:

$$\mathcal{S}\psi_{\mathcal{C}} = \mathcal{S}\kappa_{(j)}e^{(j)}\psi = \mathcal{S}\kappa_{(j)}e^{(j)}\tilde{\mathcal{S}}\mathcal{S}\psi = (\mathcal{S}\kappa_{(j)}e^{(j)}\tilde{\mathcal{S}})\mathcal{S}\psi = e^{-i\sigma_k\vartheta_k/2+x_k\alpha_k/2}\kappa_{(j)}e^{(j)}e^{i\sigma_k\vartheta_k/2-x_k\alpha_k/2}\psi' =$$

$$e^{-i\sigma_k\vartheta_k/2+x_k\alpha_k/2}\kappa_{(j)}e^{-i\sigma_{k\neq j}\vartheta_k/2+i\sigma_{(j)}\vartheta_{(j)}/2-x_{k\neq j}\alpha_k/2+x_{(j)}\alpha_{(j)}/2}e^{(j)}\psi' = \kappa_{(j)}e^{(j)}\psi'. \tag{45}$$

I illustrate this remarkable property of $\kappa_{(j)}e^{(j)}$ by showing the STR Majorana Equation (STR ME) and its Lagrangian. STR ME and the Majorana mass $m_M$ are:

$$\{i\nabla\psi = m_M\psi_{\mathcal{C}} = m_M\kappa_{(j)}e^{(j)}\psi\} \Rightarrow \{i\nabla\psi_{\mathcal{C}} = m_M\psi\} \Rightarrow \{-\nabla^2\psi = i\nabla(i\nabla\psi) = m_Mi\nabla\psi_{\mathcal{C}} = m_M^2\psi\}. \tag{46}$$

The STR Lagrangian $\mathfrak{L}_M$ is much simpler than the standard one [3] and the STR ME follows from it:

$$\mathfrak{L}_M = \overline{\psi}(i\nabla - m_M\kappa_{(j)}e^{(j)})\psi \Rightarrow \partial_\mu \delta\mathfrak{L}_M/\delta \partial_\mu\overline{\psi} - \delta\mathfrak{L}_M/\delta \overline{\psi} = -(i\nabla - m_M\kappa_{(j)}e^{(j)})\psi = 0. \tag{47}$$

$\overline{\psi}\kappa_{(j)}e^{(j)}\psi$ is the *Majorana mass term* in the STR Lagrangian. From (45) $\mathfrak{L}_M$ is obviously Lorentz-invariant.

Finally, it is straightforward to check that time reversal in (42) flips spins ($\varphi^\dagger\varphi_\mathcal{T} = \chi^\dagger\chi_\mathcal{T} = 0$ is proven in (A20) [10]). From (39, 42), three forms of $\varphi_\mathcal{T}$ are (the $j = 3$ relations below follow from the *eigen*-equations $\boldsymbol{\sigma}_3(\kappa_3 u_\varphi) = -\kappa_3 u_\varphi$ and $\boldsymbol{\sigma}_3(\kappa_3 d_\varphi) = \kappa_3 d_\varphi$):

$$\boldsymbol{\sigma}_{(j)}\kappa_{(j)}\varphi = \begin{cases} j = 1: & u_\varphi \to d'_{1\varphi} = \boldsymbol{\sigma}_1 u_\varphi^\dagger; \quad d_\varphi \to u'_{1\varphi} = -\boldsymbol{\sigma}_1 d_\varphi^\dagger \\ j = 2: & u_\varphi \to d'_{2\varphi} = i\boldsymbol{\sigma}_1 u_\varphi^\dagger; \quad d_\varphi \to u'_{2\varphi} = -i\boldsymbol{\sigma}_1 d_\varphi^\dagger \\ j = 3: & u_\varphi \to d'_{3\varphi} = -\kappa_3 u_\varphi; \quad d_\varphi \to u'_{3\varphi} = \kappa_3 d_\varphi \end{cases} \quad \begin{array}{l} u_\varphi^\dagger = \tilde{u}_\varphi^\dagger \text{ corresponds} \\ \text{to the complex conjugate} \\ \text{in the standard formalism.} \end{array} \tag{48}$$

In **conclusion**, STR promotes a geometric view of physics, where vectors and their Clifford combinations set the complex structure, not the scalar components. STR DE arises from the quantization of the classical relativistic 4-momentum vector with modulus $m$ – matrices and imposed algebra unneeded. The working of STR demonstrated in the present contribution hints to the expectation that all the formal machinery developed in nine[+] decades to handle the standard DE and its generalizations, adapts easily to the STR formalism. With its clear physical meaning and with its very efficient computation compared to matrices,

Geometric Algebra in general and STR in particular hold great potential to become standard in many areas of physics.

**Appendix**

*The STR $[x, p]$ commutator, the angular momentum $\mathbf{L}$ and the generators of the Lorentz group.*

$\mathbf{K}_j, \mathbf{J}_j$ appear as frame multivectors for the components of the Killing vectors $x_j \partial_t + t \partial_{x_j}$ and $x_j \partial_{x_k} - x_k \partial_{x_j}$ in the STR commutator $[x, p]$ of position-momentum operators. Notice by the way that $\dot{\imath}\hbar(x \wedge \nabla)$ is Lorentz invariant. The 4-position is expressed here in two equivalent forms: $x = x_\mu e^\mu = x^\mu e_\mu$, so that $x^\mu = \eta^{\mu\nu} x_\nu$:

$$[x, p] = \dot{\imath}\hbar(x\nabla - \nabla x) = \dot{\imath}\hbar(-(\nabla \cdot x) + 2x \wedge \nabla) = \dot{\imath}\hbar[-4 + 2(\mathbf{x}\partial_t + t\boldsymbol{\nabla}) + 2\dot{\imath}e^5(\mathbf{x} \times \boldsymbol{\nabla})] = -4\dot{\imath}\hbar +$$

$$2\dot{\imath}\hbar(\mathbf{x}\partial_t + t\boldsymbol{\nabla}) - 2\dot{\imath}(\boldsymbol{\sigma}, \mathbf{L}) = 4\dot{\imath}\hbar\left[-1 + \tfrac{1}{2}\mathbf{K}_j\left(x_j\partial_t + t\partial_{x_j}\right) - \tfrac{1}{2}\epsilon_{jkl}\mathbf{J}_l x_j \partial_{x_k}\right]. \quad \text{(with } \mathbf{L} = \mathbf{x} \times \mathbf{p}). \tag{A1}$$

As defined here and in Eq. (14), $\mathbf{J}_j, \mathbf{K}_j$ do not comprise $e^5$. The commutators of $\mathbf{J}_j, \mathbf{K}_j$ in STR are:

$$[\mathbf{J}_j, \mathbf{J}_k] = \epsilon_{jkl}\mathbf{J}_l; \quad [\mathbf{J}_j, \mathbf{K}_k] = \epsilon_{jkl}\mathbf{K}_l; \quad [\mathbf{K}_j, \mathbf{K}_k] = -\epsilon_{jkl}\mathbf{J}_l \quad \text{Algebra } SO(3,1) \text{ of the Lorentz group.} \tag{A2}$$

Notice the well-known similarity between (A1) and (35) for the components of the electromagnetic field ($\mathbf{L}$, ang. momentum operator). Two disjoint $SU(2)$ algebras $\mathbf{S}_{+j}, \mathbf{S}_{-j}$ emerge from the following combination of spin and boost generators, realizing the isomorphism between the algebras of $SO(3,1)$ and $SU(2) \otimes SU(2)$:

$$\mathbf{S}_{\pm j} \equiv \tfrac{1}{2}(\dot{\imath}\mathbf{J}_j \pm \mathbf{K}_j) = \tfrac{1}{2}\dot{\imath}\mathbf{J}_j(1 \pm e^5); \quad [\mathbf{S}_{+j}, \mathbf{S}_{+k}] = \epsilon_{jkl}\dot{\imath}\mathbf{S}_{+l}; \quad [\mathbf{S}_{-j}, \mathbf{S}_{-k}] = \epsilon_{jkl}\dot{\imath}\mathbf{S}_{-l}; \quad [\mathbf{S}_{+j}, \mathbf{S}_{-k}] = 0. \tag{A3}$$

In STR the Weyl left and right handed projectors $(1 \pm e^5)$ appear in (A3). Under parity $e^0 \mathbf{S}_{\pm j} e^0 = \mathbf{S}_{\mp j}$.

*Free field solutions.* A second illustration of the working of the Dirac spinors (23-26) is the solution of STR DE for the free field. As in the standard formalism, we first write the STR DE as two coupled equations. This form is also used further down in the derivation of the STR Pauli Equation. With the STR Dirac spinor $\psi = \varphi + \chi$ from (23), we write down the two equations obtained by the sum and difference of the STR DE (11) and the parity-transformed STR DE (below I use the shorthand $P_\mu = \dot{\imath}\hbar\partial_\mu + eA_\mu$ and $\mathbf{P} = P_j \mathbf{x}_j$):

$$\begin{cases}(P_0 e^0 + P_j e^j - m)(\varphi + \chi) = 0 \\ (P_0 e^0 - P_j e^j - m)(\varphi - \chi) = 0\end{cases} \Rightarrow \begin{cases}(P_0 e^0 - m)\varphi + P_j e^j \chi = 0 \\ (P_0 e^0 - m)\chi + P_j e^j \varphi = 0\end{cases} \Rightarrow \begin{cases}(P_0 - m)\varphi - \mathbf{P}\chi = 0 \\ (P_0 + m)\chi - \mathbf{P}\varphi = 0.\end{cases} \tag{A4}$$

In the last step we use $P_0 e^0 \varphi = P_0 \varphi$; $P_0 e^0 \chi = -P_0 \chi$, in accordance with the definition of $\varphi, \chi$ in (23) and the effect of parity in (24). The other piece of preparation we need is that again as in the standard case, the free field STR DE spinor can be expanded in plane waves of positive and negative energy and a constant spinor depending only on the 4-momentum p and the two spin degrees of freedom $s$ of the free particle:

$$\psi_+ = e^{-ip\cdot x/\hbar}u(p,s) \quad \text{and} \quad \psi_- = e^{ip\cdot x/\hbar}v(p,s); \qquad u(p,s), v(p,s) \text{ satisfy STR DE}. \tag{A5}$$

Therefore, $u(p,s)$ and $v(p,s)$, similarly to $\psi$ in (23) can be expressed as pairs of Pauli spinors:

$$u(p,s) = \varphi_+ + \chi_+; \quad v(p,s) = \varphi_- + \chi_- \quad \text{with} \quad \varphi_+, e^5\chi_+, e^5\varphi_-, \chi_- \in \Sigma. \tag{A6}$$

Then the form of the STR DE as two coupled equations in (A4) applies for each pair of spinors in (A6):

$$\begin{cases}(E-m)\varphi_+ - \mathbf{p}\chi_+ = 0 \\ (E+m)\chi_+ - \mathbf{p}\varphi_+ = 0\end{cases} \quad \text{and} \quad \begin{cases}(E-m)\varphi_- - \mathbf{p}\chi_- = 0 \\ (E+m)\chi_- - \mathbf{p}\varphi_- = 0.\end{cases} \tag{A7}$$

$E = p_0$ is the energy (scalar) and $\mathbf{p}$ is the 3-momentum (vector in $\mathbf{X}$ from (13)); the momentum of the free particle being a constant of the motion, we can use the vector instead of the operator. The first $\varphi$-terms in the upper equations belong to $\Sigma$ and so do the second $\chi$-terms, e.g. $\mathbf{p}\chi = \mathbf{p}e^5e^5\chi = p_k\sigma_k e^5\chi \equiv (\mathbf{p},\boldsymbol{\sigma})(e^5\chi) \in \Sigma$ (see (14)). Similarly, left-multiplication by $e^5$ of the lower equations in (A7) brings all their terms into $\Sigma$. After this consistency check, we can proceed with the free field solutions. For positive energy $E > 0$, the factor $E + m > 0$, therefore we can express $\chi_+$ from the lower equation in (A7) as a function of $\varphi_+$ and then substitute it into the upper equation:

$$\chi_+ = \mathbf{p}\varphi_+/(E+m); \quad (E^2 - m^2 - \mathbf{p}^2)\varphi_+ = 0. \tag{A8}$$

For $\varphi_+ \neq 0$ the last Equation is just the relativistic invariant $E^2 - m^2 - \mathbf{p}^2 = 0$. As shown above, we can write the first Equation in (A7) as $e^5\chi_+ = (\mathbf{p},\boldsymbol{\sigma})\varphi_+/(E+m)$. Now we can express $e^5\chi_+, \varphi_+$ by the corresponding probability amplitudes for spin up and spin down in Eq. (25):

$$e^5\chi_+ = u_{\chi+} + d_{\chi+} = (\mathbf{p},\boldsymbol{\sigma})(u_{\varphi+} + d_{\varphi+})/(E+m) \quad \text{with} \quad u_{\varphi+}, \sigma_1 d_{\varphi+}, u_{\chi+}, \sigma_1 d_{\chi+} \in \{a+b\mathbf{i}\}. \tag{A9}$$

The last equation is easily solved:

$$u_{\chi+} = [p_3 u_{\varphi+} + (p_1 - \mathbf{i}p_2)\sigma_1 d_{\varphi+}]/(E+m); \quad d_{\chi+} = [-p_3 d_{\varphi+} + (p_1 + \mathbf{i}p_2)\sigma_1 u_{\varphi+}]/(E+m). \tag{A10}$$

With the plane wave prefactor in (A5), the general solution (non normalized) takes in STR the form:

$$E > 0: \psi_+ = e^{-ip\cdot x/\hbar}\left[u_{\varphi+} + d_{\varphi+} + e^5\left(\frac{p_3 u_{\varphi+} + (p_1 - \mathbf{i}p_2)\sigma_1 d_{\varphi+}}{E+m}\right) + \sigma_1\frac{-p_3\sigma_1 d_{\varphi+} + (p_1 + \mathbf{i}p_2)u_{\varphi+}}{E+m}\right] \in \Xi, \tag{A11}$$

which in the case of spin up (↑) ($u_{\varphi+} = 1$), respectively spin down (↓) ($\sigma_1 d_{\varphi+} = 1$) yields:

$$\psi_{+\uparrow} = e^{-ip\cdot x/\hbar}\left[1 + e^5\left(\frac{p_3}{E+m} + \sigma_1\frac{p_1+\mathbf{i}p_2}{E+m}\right)\right]; \quad \psi_{+\downarrow} = e^{-ip\cdot x/\hbar}\left[\sigma_1 + e^5\left(\frac{p_1-\mathbf{i}p_2}{E+m} - \sigma_1\frac{p_3}{E+m}\right)\right] \tag{A12}$$

In the rest frame of the electron $p = me_{rf}^0$ so that (A11) reduces to the explicit solution to Eq. (29):

Rest frame ($\tau$ proper time). $E = m > 0$: $\psi_{+r} = e^{-\mathrm{i}m\tau/\hbar}\varphi_{+rf} = e^{-\mathrm{i}m\tau/\hbar}(u_{\varphi+} + d_{\varphi+})_{rf}.$ \hfill (A13)

We will meet the 'fast oscillations' factor $e^{-\mathrm{i}mt/\hbar}$ for slow electrons ($t \approx \tau$) in Eq. (A18) leading to the STR Pauli Equation (A19). Similarly to (A11) one finds the 'negative energy' solutions from the second pair of equations in (A7), in this case recalling that $-E + m > 0$. We just show the result for $E < 0$ (see (A6)):

$$\psi_- = \varphi_- + \chi_- = e^{\mathrm{i}p \cdot x/\hbar}\left[e^5\left(\frac{-p_3 u_{\chi-} - (p_1 - \mathrm{i}p_2)\sigma_1 d_{\chi-}}{-E+m}\right) + \sigma_1\frac{p_3\sigma_1 d_{\chi-} - (p_1 + \mathrm{i}p_2)u_{\chi-}}{-E+m}\right) + u_{\chi-} + d_{\chi-}\right] \in \Xi. \quad (A14)$$

As expected, the general $\psi_\pm$ in (A11, A14) are part of the real subalgebra $\Xi \subset X$ with basis shown in (13).

I conclude this section by a short discussion of Lorentz transformations of the free field $\psi_+$. From (A5, A6):

$$\mathcal{L}: \{\psi_+ = e^{-\mathrm{i}p \cdot x/\hbar} u(p, s)\} \to \{\psi'_+ = \mathcal{S}\psi_+ = Se^{-\mathrm{i}p_\mu x^\mu/\hbar}\tilde{S}Su(p, s) = e^{-\mathrm{i}p'_\mu x'^\mu/\hbar}S(\varphi_+ + \chi_+)\} \quad (A15)$$

The scalar product $p \cdot x$ is Lorentz invariant, i.e. $p \cdot x = p' \cdot x' = S(p \cdot x)\tilde{S} = \frac{1}{2}S(px + xp)\tilde{S}$; in components $p_\mu x^\mu = p'_\mu x'^\mu$. Looking at the last term in (A15), we render explicit the spin components from (25):

$$S(\varphi_+ + \chi_+) = S[u_{\varphi+} + d_{\varphi+} + e^5(u_{\chi+} + d_{\chi+})] = S(u_{\varphi+} + d_{\varphi+}) + e^5 S(u_{\chi+} + d_{\chi+}). \quad (A16)$$

(A16) illustrates that at the level of Pauli spinors, $\mathcal{S}$ reduces to a one-sided S Lorentz transform. However, as stated in (A5, A6) the spinor $u(p, s)$ depends on the momentum $p$, therefore the general transformation in (A16) should be $\mathcal{S}$, not S. This is indeed the case at the solution level, where the operator part 'delivers' the momentum components of $u(p, s)$, as one can see in Eq. (A9) and more explicitly in the solutions $\psi_\pm$, Eqs. (A11, A14). Concentrating on the $\psi_+$ case, I illustrate the action of $\mathcal{S}$ on $u(p, s) = \varphi_+ + \chi_+$ in the case of a rotation in a plane normal to $\sigma_2$ by an angle of $\vartheta$. In Eq. (A9) $u_{\chi+} + d_{\chi+}$ is expressed in terms of $u_{\varphi+} + d_{\varphi+}$ and the 4-momentum components from the operator part in the form $\frac{(\mathbf{p},\sigma)}{E+m} = \frac{p_j \sigma_j}{E+m}$. As in Eq. (A16) the spinors transform by a one-sided S operator (for simplicity of notation, I remove the + subscripts):

$$S(u_\varphi + d_\varphi) = e^{J_2\vartheta/2}(u_\varphi + d_\varphi) = \left(\cos\tfrac{\vartheta}{2} - \mathrm{i}\sigma_2\sin\tfrac{\vartheta}{2}\right)(u_\varphi + d_\varphi) = \left(u_\varphi\cos\tfrac{\vartheta}{2} - \sigma_1 d_\varphi\sin\tfrac{\vartheta}{2}\right) +$$

$\left(d_\varphi\cos\tfrac{\vartheta}{2} + \sigma_1 u_\varphi\sin\tfrac{\vartheta}{2}\right) = u'_\varphi + d'_\varphi$. Notice the characteristic half angle for spinor transformation.

$$S\chi = e^5(u'_\chi + d'_\chi) = e^5 S\frac{(\mathbf{p},\sigma)}{E+m}(u_\varphi + d_\varphi) = e^5 S\frac{(\mathbf{p},\sigma)}{E+m}\tilde{S}S(u_\varphi + d_\varphi) = e^5\frac{(\mathbf{p}',\sigma')}{E'+m}(u'_\varphi + d'_\varphi), \quad \text{where:}$$

$(\mathbf{p}, \sigma) = p_j\sigma_j$;  $p'_\mu = e'_\mu \cdot p = \mathrm{Re}_\mu \tilde{R} \cdot p$;  $e'_1 = (\cos\vartheta - \mathrm{i}\sigma_2\sin\vartheta)e_1 = e_1\cos\vartheta - e_3\sin\vartheta$;  $p'_1 = p_1\cos\vartheta - p_3\sin\vartheta$;  $e'_2 = e_2$;  $p'_2 = p_2$;  $e'_3 = e_3\cos\vartheta + e_1\sin\vartheta$;  $p'_3 = p_3\cos\vartheta + p_1\sin\vartheta$;  $e'_0 = e_0$;  $E' = E$;  $\sigma'_1 = R\sigma_1\tilde{R} = \sigma_1\cos\vartheta - \sigma_3\sin\vartheta$;  $\sigma'_2 = \sigma_2$;  $\sigma'_3 = R\sigma_3\tilde{R} = \sigma_3\cos\vartheta + \sigma_1\sin\vartheta$. \hfill (A17)

***The STR Pauli Equation, STR PE.*** The lowest order nonrelativistic approximation to STR DE (A4) yields the STR PE. Following Feynman (see also (A13)), we isolate the fast oscillating part of $\psi$ as a common factor $\rho = \rho(t)$ to $\varphi$ and $\chi$ in (A4), leaving behind the nonrelativistic Pauli spinors proper $\varphi_P, \chi_P$:

$$\begin{cases}(P_0 - m)\rho\varphi_P - \mathbf{P}\rho\chi_P = 0 \\ (P_0 + m)\rho\chi_P - \mathbf{P}\rho\varphi_P = 0\end{cases} \xRightarrow{\rho=e^{-\mathrm{i}mt/\hbar}} \begin{cases}(\mathrm{i}\hbar\partial_t + eA_0)\varphi_P - \mathbf{P}\chi_P = 0 \\ (\mathrm{i}\hbar\partial_t + eA_0 + 2m)\chi_P - \mathbf{P}\varphi_P = 0.\end{cases} \quad (A18)$$

For $|(\mathrm{i}\hbar\partial_t + eA_0)\chi_P| \ll 2m|\chi_P|$ (nonrelativistic regime) the lower equation approximates in lowest order to: $\chi_P \approx \mathbf{P}\varphi_P/2m$. I.e. for slow electrons $|\chi_P| \ll |\varphi_P|$. Substituting into the upper equation one obtains the Pauli Hamiltonian [14] $H_P$ (below: $\mathbf{PP} = \mathbf{P} \cdot \mathbf{P} + \mathbf{P} \wedge \mathbf{P} = \mathbf{P} \cdot \mathbf{P} + \hbar e(\boldsymbol{\sigma}, \mathbf{B})$, where $\mathbf{P} \cdot \mathbf{P} = (\mathbf{p} + e\mathbf{A}) \cdot (\mathbf{p} + e\mathbf{A}) = (-\hbar\mathrm{i}\nabla + e\mathbf{A}) \cdot (-\hbar\mathrm{i}\nabla + e\mathbf{A}) \equiv \mathbf{P}^2$ is a grade 0 + 5 operator):

$$\mathrm{i}\hbar\partial_t\varphi_P = H_P\varphi_P = \left[\frac{\mathbf{P}^2}{2m} - eA_0 + \frac{\hbar e}{2m}(\boldsymbol{\sigma}, \mathbf{B})\right]\varphi_P. \quad (A19)$$

This is the STR Pauli Equation (PE), identical in form to the standard PE [14], but here without matrices and with a complex structure surging from the real vector space STR! The term $(\hbar e/2m)(\boldsymbol{\sigma}, \mathbf{B})$ mentioned in the main text, marks the additional potential energy due to the spin magnetic moment of a slow electron. It distinguishes STR PE from the STR Schrödinger Equation [15], which is obtained from (A19) by removing it (no spin) and by freezing the spinor $\varphi$ at spin up.

***Orthogonality between $\psi$ and $\psi_{\mathcal{T}}$***, i.e. $\overline{\psi}\psi_{\mathcal{T}} = 0$ illustrated for the case $j = 3$, Eq. (43). In all three cases:

$$\overline{\psi}\psi_{\mathcal{T}} = \varphi^\dagger\varphi_{\mathcal{T}} - \chi^\dagger\chi_{\mathcal{T}} = \varphi^\dagger\boldsymbol{\sigma}_{(j)}\kappa_{(j)}\varphi - \chi^\dagger\boldsymbol{\sigma}_{(j)}\kappa_{(j)}\chi. \quad \text{Looking now at the case } j = 3:$$

$$\varphi^\dagger\boldsymbol{\sigma}_3\kappa_3\varphi = \left(u_\varphi^\dagger + d_\varphi^\dagger\right)\left(u'_{3\varphi} + d'_{3\varphi}\right) = u_\varphi^\dagger\kappa_3 d_\varphi - d_\varphi^\dagger\kappa_3 u_\varphi = u_\varphi^\dagger d_\varphi^\dagger - d_\varphi^\dagger u_\varphi^\dagger = 0. \quad (A20)$$

In the same way $\chi^\dagger\boldsymbol{\sigma}_3\kappa_3\chi = 0$, which completes the proof. For all the three cases $j = 1,2,3$ in Eq. (43) one can show similarly that $\overline{\psi}\psi_{\mathcal{T}} = \alpha_j\left[\left(u_\varphi^\dagger d_\varphi^\dagger - d_\varphi^\dagger u_\varphi^\dagger\right) - \left(u_\chi^\dagger d_\chi^\dagger - d_\chi^\dagger u_\chi^\dagger\right)\right] = 0; \alpha_1 = -\boldsymbol{\sigma}_1, \alpha_2 = -\mathrm{i}\boldsymbol{\sigma}_1, \alpha_3 = 1.$